\documentclass[pra,twocolumn,aps,superscriptaddress,longbibliography,notitlepage,floatfix]{revtex4-1}

\usepackage{amsmath}
\usepackage{amssymb}
\usepackage{mathrsfs}
\usepackage{graphicx}
\usepackage{float}
\usepackage[caption=false]{subfig}
\usepackage{epstopdf}
\usepackage{xcolor}

\usepackage[colorlinks,linkcolor={red!75!black},citecolor={red!75!black},urlcolor={blue!75!black}]{hyperref}

\definecolor{brightmaroon}{rgb}{0.76,0.13,0.28}

\renewcommand{\Re}{\operatorname{Re}}

\newcommand{\Tr}{\mathrm{Tr}}
\newcommand{\Rnum}[1]{\uppercase\expandafter{\romannumeral #1\relax}}

\def\Chen#1{#1}
\def\QY#1{#1}

\def\CH#1{#1}

\begin{document}

\title{Quantum Mpemba effect of \texorpdfstring{$\mathbb{Z}_2$}{Z2} symmetry restoration in the Dicke model}

\author{Dongni Chen}
\affiliation{Research Center for Advanced Optics and Photoelectronics, Department of Physics, College of Science, Shantou University, Shantou 515063, China}

\author{Wei Xiong}
\affiliation{Research Center for Advanced Optics and Photoelectronics, Department of Physics, College of Science, Shantou University, Shantou 515063, China}

\author{Peng-Cheng Li}
\affiliation{Research Center for Advanced Optics and Photoelectronics, Department of Physics, College of Science, Shantou University, Shantou 515063, China}

\author{Yee Sin Ang}
\affiliation{Science, Mathematics and Technology, Singapore University of Technology and Design, Singapore 487372, Singapore}

\author{Ching Hua Lee}\email{phylch@nus.edu.sg}
\affiliation{Department of Physics, National University of Singapore, Singapore 117542}

\author{Yi Qin}\email{doctor\_yi\_qin\_25@nus.edu.sg} 
\affiliation{Department of Physics, National University of Singapore, Singapore 117542}

\begin{abstract}
We study a symmetry-restoration quantum Mpemba effect in the dissipative Dicke model as an example of a weakly $\mathbb Z_2$-symmetric open quantum system, where the full Liouvillian is parity covariant although individual quantum jumps connect the two Hilbert-space parity sectors. Using the entanglement-asymmetry (EA) diagnostic to monitor parity restoration, we compare coherent-product and \textcolor{black}{cat-product} initial states. Coherent-product states show conventional relaxation, with larger initial asymmetry decaying more slowly and no Mpemba crossing observed. \textcolor{black}{Cat-product states}, by contrast, can reverse this hierarchy over a finite region of initial-state space. This reversal is enabled by initial-state mode filtering: the \textcolor{black}{cat-product superposition} suppresses the contributions of the leading slow Liouvillian modes, allowing a more asymmetric state to restore parity through faster decay channels. Our results reveal that symmetry-restoration QME can arise even when parity is preserved only as a weak Liouvillian symmetry, with quantum jumps mixing the Hilbert-space parity sectors. The anomalous relaxation is governed by initial-state-dependent filtering of the leading slow Liouvillian modes, providing a mechanism for symmetry-restoration QME beyond strongly conserved symmetry sectors.
\end{abstract}

\maketitle

\section{Introduction}
\label{sec:intro}

The Mpemba effect refers to an anomalous relaxation phenomenon in which a state initially farther from equilibrium reaches the steady state faster than one initially closer to it. Since its early observation in the freezing dynamics of water \cite{MpembaOsborne1969}, it has been generalized to a broad class of classical and nonequilibrium systems \cite{BurridgeLinden2016,Lasanta2017GranularPRL,Tian2025PRA,Tang2023DirectObservation}. Its quantum counterpart, the quantum Mpemba effect (QME), has recently attracted growing interest in many-body and open-system dynamics \cite{Ares2025NatureReviewsPhysics,Teza2026PhysicsReports,Yu2025QuantumMpembaSymmetry,Turkeshi2025PRL,Kochsiek2022PRA}, with manifestations reported in dissipative systems~\cite{Longhi2025NonNormal,Longhi2025APLQuantum,Zhao2026CommunicationsPhysics,Longhi2026Entropy,Furtado2025AnnalsofPhysics,Nava2024PRL,Chatterjee2023PRL,Santos2026SuperradiantMpemba}, non-Markovian dynamics~\cite{Strachan2025NonMarkovPRL}, integrable and nonintegrable models~\cite{Ares2025PRB_FreeFermionicMixed}, long-range interacting
systems~\cite{Yamashika2026PRL}, symmetry-resolved relaxation \cite{Yu2025QuantumMpembaSymmetry,Li2026PRB,Summer2026PRX}, non-Hermitian systems~\cite{Ma2025PRB}, and imaginary-time dynamics \cite{Chang2026PRL,Yu2026PRB}. For open quantum systems, this phenomenon is naturally formulated in terms of the Liouvillian spectrum, where the approach to the steady state is controlled by decay modes with different decay rates \cite{GoriniKossakowskiSudarshan1976,Lindblad1976,BreuerPetruccione2002,GardinerZoller2004,Carmichael1993,Macieszczak2016PRL}. Within this framework, a microscopic route to the QME arises when the initial state has a suppressed overlap with the slowest non-stationary mode, so that faster subleading modes dominate the observable relaxation \cite{CarolloLasantaLesanovsky2021,Zhou2023PRR_LEP,MoroderGoold2024,wei2026quantumScienceChinaPhysics,Zhang2025StrongMpemba}.

A particularly useful route to the QME in many-body systems is provided by exact conservation laws, such as $U(1)$ charge conservation. \CH{Conservation laws can generate distinct symmetry-resolved relaxation pathways with different characteristic timescales, thereby providing additional mechanisms for anomalous relaxation ordering \cite{Liu2024PRLcircuits,liu2025ScienceBulletin,Rylands2024PRL}.}
This perspective naturally raises the question of how symmetry breaking and restoration should be quantified during the relaxation process. 
Entanglement asymmetry \CH{(EA)} provides an operational diagnostic for this purpose. It compares a density matrix with its sector-projected counterpart and thereby quantifies the degree of symmetry breaking \cite{AresMurcianoCalabrese2023,Joshi2024ObservingQME,Fujimura2026NonAbelianEA,Caceffo2024JournalofStatisticalMechanics,Yamashika2024PRB}, allowing one to track symmetry restoration during nonequilibrium relaxation.


Yet, symmetry-restoration QME in open systems with only a weak discrete $\mathbb{Z}_2$ symmetry represents a
distinct and largely unexplored setting. Here weak symmetry means covariance of the full Liouvillian under parity conjugation, rather than conservation of the even and odd Hilbert-space parity sectors by every quantum jump. This distinguishes weakly
$\mathbb{Z}_2$-symmetric dynamics from conventional symmetry-restoration scenarios, where strong conservation laws or
extensive symmetry-sector structures provide well-separated relaxation pathways. Recent studies have shown that quantum Mpemba behavior can nevertheless emerge without global symmetries \cite{Bhore2025PRB}, in measurement-induced dynamics
\cite{DiGiulio2025MeasurementInduced}, and in other symmetry-restoration settings
\cite{Chalas2024JournalofStatisticalMechanics, Yu2025ChinesePhysicsLetters}. However, it remains unclear how
anomalous symmetry-restoration dynamics can arise in weakly $\mathbb{Z}_2$-symmetric Lindblad systems, where quantum jumps mix the two parity sectors and relaxation is governed by Liouvillian decay modes rather than independently conserved symmetry sectors. In particular, whether the initial state can selectively suppress the dominant slow Liouvillian contribution and thereby reverse the parity-restoration hierarchy remains an open question.

In this work, we address this question in the dissipative Dicke model, a paradigmatic light--matter system with a discrete parity symmetry \cite{Dicke1954,WangHioe1973,HeppLieb1973,EmaryBrandes2003,Ritsch2013,KirtonKeeling2019Review}. 
The Dicke model describes the collective coupling between a single bosonic mode and an ensemble of two-level atoms, and exhibits a well-known normal-to-superradiant transition accompanied by spontaneous $\mathbb{Z}_2$ symmetry breaking in the thermodynamic limit \cite{WangHioe1973,HeppLieb1973,EmaryBrandes2003}. 
In experimentally relevant realizations, dissipation is unavoidable and gives rise to open system dynamics governed by a Liouvillian superoperator
\cite{Dimer2007,Baumann2010Nature,Nagy2011OpenDicke,KirtonKeeling2017PRL,GelhausenBuchhold2018PRA,FornDiaz2019USC_RMP,Blais2021CircuitQED_RMP}. Importantly, cavity photon loss, as well as collective atomic decay when included, connects the two Hilbert-space parity sectors, while the full Liouvillian \textcolor{black}{remains covariant under parity conjugation}. The open Dicke model therefore possesses a weak $\mathbb{Z}_2$ symmetry rather than strong parity conservation, making it a natural platform for defining and testing symmetry-restoration QME under weakly symmetric dissipative dynamics.

To uncover the mechanisms underlying QME behavior in the open Dicke model, we investigate the interplay between initial-state structure and the Liouvillian spectrum. We compare two representative classes of initial states. 
The first consists of coherent-product states, whose parity breaking is mainly encoded in a classical-like cavity displacement and whose cavity component is an eigenstate of the photon-loss jump operator. 
The second consists of \textcolor{black}{cat-product states}, which coherently combine two parity-related coherent states. 
To characterize parity restoration during the dissipative evolution, we use the EA diagnostic defined in Sec.~\ref{sec:model} and complement it with a direct decomposition into Liouvillian eigenmodes.

We show that coherent-product states exhibit conventional relaxation behavior, with no crossings between EA trajectories. 
By contrast, \textcolor{black}{cat-product states} display clear and robust crossings, signaling a QME.
We identify the underlying mechanism as an initial-state mode-filtering effect, in which the coherent superposition of two parity-related coherent states suppresses the projections onto the leading slow Liouvillian modes and allows faster modes to dominate the relaxation dynamics.
Across the normal and superradiant regimes, the same filtering mechanism remains operative and gives rise to Mpemba crossings. In both cases the EA decays monotonically, but the low-lying Liouvillian structure determines whether the relaxation develops a pronounced step-like profile or remains comparatively smooth without a pronounced plateau. These results establish symmetry-restoration QME under weak $\mathbb{Z}_2$ Lindblad symmetry and identify initial-state mode filtering as the mechanism that suppresses the contributions of the leading slow Liouvillian modes and reverses the relaxation ordering.

The remainder of this paper is organized as follows. 
Section~II introduces the dissipative Dicke model and its $\mathbb{Z}_2$ symmetry structure. 
Section~III analyzes the relaxation dynamics for different initial states, identifies the initial-state mode-filtering mechanism behind the Mpemba crossings, and discusses how the low-lying Liouvillian structure shapes the detailed EA relaxation profile.
Section~IV maps out the parameter-space structure of the anomalous relaxation behavior.
Finally, Sec.~V summarizes our main results and discusses possible experimental implications.

\section{Open Dicke model, weak parity symmetry, and the entanglement asymmetry diagnostic}
\label{sec:model}


The QME studied here concerns $\mathbb Z_2$ symmetry restoration in a weakly $\mathbb Z_2$-symmetric open system. In this setting, the full Lindblad evolution respects parity covariance, whereas individual quantum jumps need not conserve Hilbert-space parity. Consequently, the even and odd Hilbert-space parity sectors are not independently conserved. The central question is therefore whether anomalous relaxation ordering can still emerge in such a weak-symmetry setting and, if so, how it is encoded in the initial-state projection onto the Liouvillian decay modes. To address this question with minimal ingredients, we consider the open Dicke model, which describes a single bosonic mode collectively coupled to an ensemble of $N$ identical two-level atoms. \CH{Its non-dissipative processes are described by the Hamiltonian}
\begin{equation}
H = \omega a^{\dagger} a + \omega_0 J_z
+ \frac{\lambda}{\sqrt{N}} (a + a^{\dagger})(J_+ + J_-),
\label{Eq:Hamiltonian}
\end{equation}
where $a^{\dagger}$ and $a$ are the creation and annihilation operators of the cavity mode with frequency $\omega$, and
\begin{equation}
J_z=\frac{1}{2}\sum_{i=1}^{N}\sigma_z^{i},
\qquad
J_{\pm}=\sum_{i=1}^{N}\sigma_{\pm}^{i}
\end{equation}
are the collective spin operators for the atomic ensemble, where $\sigma_\pm^i=(\sigma_x^i\pm i\sigma_y^i)/2$. \textcolor{black}{This dissipative Dicke model admits implementations in trapped-ion~\cite{Genway2014PRL,Gambetta2019PRA} and circuit-QED~\cite{Felicetti2018PRA} platforms. Beyond these settings, programmable quantum processors have also demonstrated capabilities for investigating nonequilibrium many-body dynamics~\cite{2022Mi-nature,2022Koh-PRL} and symmetry-related quantum phenomena~\cite{Shen2025npjQuantumInformation,shen2026resolving,2021Satzinger-Science,2021Semeghini-Science}.}

The Hamiltonian is invariant under the discrete parity transformation generated by
\begin{equation}
\hat{P}=\exp\!\left[i\pi\left(a^{\dagger}a+J_z+j\right)\right],
\label{Eq:excitations}
\end{equation}
with $j=N/2$, where $\hat P$ measures the parity of the total excitation number, including both cavity photons and atomic excitations. Its action on the elementary operators is

\begin{equation}
\hat{P} a \hat{P}^{\dagger}=-a,
\qquad
\hat{P} J_{\pm} \hat{P}^{\dagger}=-J_{\pm},
\end{equation}
and therefore
\begin{equation}
[H,\hat{P}]=0.
\end{equation}
To incorporate dissipation, we study the open-system dynamics within the Lindblad master-equation formalism,
\begin{equation}
\frac{d\rho}{dt}
= -i[H,\rho]
+ \sum_{\mu}
\left(
L_{\mu}\rho L_{\mu}^{\dagger}
-\frac{1}{2}\{L_{\mu}^{\dagger}L_{\mu},\rho\}
\right).
\label{eq:lindblad}
\end{equation}
We consider cavity photon loss described by the jump operator $L_\kappa=\sqrt{\kappa}a$, neglecting atomic decay.

{\color{black}Although photon loss mixes the Hilbert-space parity sectors, the Liouvillian remains covariant under parity conjugation:}
\begin{equation}
\mathcal{L}\!\left[\hat P\rho\hat P^\dagger\right]
=
\hat P\,\mathcal{L}[\rho]\,\hat P^\dagger.
\end{equation}
Introducing the parity-conjugation superoperator
\begin{equation}
\mathcal{P}[\rho]=\hat P\rho\hat P^\dagger,    
\end{equation}
this relation becomes
\begin{equation}
[\mathcal{L},\mathcal{P}]=0 .
\end{equation}
\QY{Thus, the open Dicke dynamics possesses only weak $\mathbb{Z}_2$ symmetry: while the photon-loss jump operator $a$ connects the two Hilbert-space parity sectors, the full Liouvillian \textcolor{black}{commutes with the parity-conjugation superoperator}.} Equivalently, $a$ is odd under parity, but the dissipator is unchanged because $\mathcal D[-a]=\mathcal D[a]$. This realizes the weak-symmetry setting introduced above: parity is not conserved by individual quantum jumps, but parity conjugation remains an exact symmetry of the full Lindblad evolution. The dissipative Dicke model therefore exhibits distinct physics from both the closed Dicke model and a strongly symmetric Lindblad system. In the closed case, the dynamics is unitary and the normal-to-superradiant transition is characterized by the Hamiltonian spectrum and ground-state symmetry breaking. In its dissipative weak-symmetry setting, the approach to the stationary state is instead governed by the Liouvillian spectrum. Together with collective light--matter coupling, this makes parity restoration sensitive to both the structure of low-lying decay modes and the coherence structure of the initial state at the level of the density matrix, which determines its overlap with different Liouvillian modes.


\QY{Our setup exhibits a dissipative normal-to-superradiant transition at a mean-field critical coupling $\lambda_c$, which gives rise to two distinct relaxation regimes in which the consequences of the weak $\mathbb{Z}_2$ symmetry can be examined.} For cavity photon loss $L_\kappa=\sqrt{\kappa}a$, the mean-field critical coupling is
\begin{equation}
\lambda_c
=
\frac{1}{2}
\sqrt{
\frac{\left(\omega^2+\kappa^2/4\right)\omega_0}{\omega}
}. 
\end{equation}
\textcolor{black}{This result follows from a linear-stability analysis of the normal state, whose leading fluctuation eigenvalue vanishes at $\lambda_c$ in the $4\times4$ Holstein--Primakoff drift matrix~\cite{2003EmaryPRL,EmaryBrandes2003} (see Appendix~\ref{app:linear_fluctuation}).}
For $\lambda<\lambda_c$, the system is in the normal phase, characterized by a vanishing cavity displacement and a polarized spin ensemble. For $\lambda>\lambda_c$, it enters the superradiant phase, where the cavity mode acquires a macroscopic occupation and parity is spontaneously broken at the mean-field level in the thermodynamic limit. In the following, we use representative values of $\lambda$ on the two sides of this dissipative transition to compare parity-restoration dynamics in the normal and superradiant regimes. The normal and superradiant regimes exhibit distinct Liouvillian spectral structures, allowing us to examine how initial-state mode filtering shapes QME dynamics in each regime.

To quantify the QME in this weak-symmetry setting, we need an observable measure of parity restoration. We use the second-R\'enyi parity asymmetry, i.e. the full-density-matrix analogue of the entanglement asymmetry (EA) introduced in Ref.~\cite{AresMurcianoCalabrese2023}. \CH{Hereafter, this will be the only version of the EA that is used. Since the states generated by the Lindblad dynamics are generally mixed, we apply this construction directly to the full density matrix.} 
The EA is defined as
\begin{equation}
\mathrm{EA}(t)\equiv \Delta S^{(2)}(t)
=
S^{(2)}\!\left[\rho_{P}(t)\right]
-
S^{(2)}\!\left[\rho(t)\right],
\label{eq:EA_Renyi2}
\end{equation}
where  $S^{(2)}(\rho)=-\ln \Tr(\rho^2)$ is the second-R\'enyi entropy, and $\rho_{P}(t)$ is the symmetry-projected density matrix 
\begin{equation}
\rho_{P}(t)=\sum_{p=\pm 1}\Pi_p\,\rho(t)\,\Pi_p,
\quad
\Pi_p=\frac{1}{2}\bigl(1+p\,\hat P\bigr),
\end{equation}
\CH{where $\Pi_+$ and $\Pi_-$ project onto the even and odd parity sectors of the full Dicke Hilbert space, respectively, with $\hat P$ defined in Eq.~\ref{Eq:excitations}. In all, the EA $\Delta S^{(2)}(t)$ compares the density matrix with its parity-projected counterpart and thereby measures the parity-sector coherence removed by the projection with respect to the global parity symmetry of the full Dicke system.}

{\color{black}
The parity-asymmetry diagnostic admits an equivalent representation in Liouville space. Introducing
\begin{equation}
\rho_\pm(t)=\frac{1}{2}\left[\rho(t)\pm\hat P\rho(t)\hat P^\dagger\right],
\label{eq:rho_parity_components}
\end{equation}
we have
\begin{equation}
\rho_P(t)=\rho_+(t),
\qquad
\Tr\!\left[\rho_+(t)\rho_-(t)\right]=0.
\end{equation}
Consequently,
\begin{equation}
\mathrm{EA}(t)=
\ln\!\left[
1+\frac{\|\rho_-(t)\|_{\rm HS}^2}
{\|\rho_+(t)\|_{\rm HS}^2}
\right],
\label{eq:EA_parity_odd}
\end{equation}
where $\|A\|_{\rm HS}^2=\Tr(A^\dagger A)$. Since the Liouvillian commutes with parity conjugation, the parity-even and parity-odd operator sectors evolve independently. The restoration of parity symmetry is therefore governed by the decay of the parity-odd component of the density matrix.
}


A \CH{non-vanishing} $\mathrm{EA}(t)$ signals residual coherence between the even and odd parity sectors of the density matrix, while its decay tracks the restoration of parity invariance. We therefore define a symmetry-restoration QME through an inversion of the EA relaxation ordering. For two initial states $\rho_0^{(1)}$ and $\rho_0^{(2)}$, the QME occurs if
\begin{equation}
\begin{aligned}
&\exists\, t_c > 0 \quad \text{such that}\\
&\mathrm{EA}_1(t) > \mathrm{EA}_2(t),
&& 0 \leq t < t_c,\\
&\mathrm{EA}_1(t_c) = \mathrm{EA}_2(t_c),\\
&\mathrm{EA}_1(t) < \mathrm{EA}_2(t),
&& \forall\, t > t_c.
\end{aligned}
\end{equation}
\CH{The QME is present, i.e., when the state with greater initial parity asymmetry ``overtakes" the relaxation of the other state and ends up with restored weak $\mathbb Z_2$ symmetry sooner.} The crossing of the two EA curves is the operational signature of QME used below. The analysis below applies this diagnostic to coherent-product states and \textcolor{black}{cat-product states}. As shown in the following section, their distinct projections onto the slow Liouvillian modes lead to qualitatively different relaxation orderings and hence QME prospects.

\section{Quantum Mpemba effect in the open Dicke model}
\label{sec:QME}
We first investigate whether weakly $\mathbb{Z}_2$-symmetric open dynamics can exhibit a QME and how the relaxation ordering depends on the structure of the initial state. \Chen{We compare two representative
classes of parity-structured states: coherent-product states [Eq.~\ref{eq:coherent_state}], where the parity asymmetry is associated with a classical-like cavity displacement, and  cat-product states [Eq.~\ref{eq:cat_state}], formed by a superposition of two parity-related cavity coherent states.  In both cases, the atomic ensemble is initialized in $|\downarrow\rangle^{\otimes N}$. The coherent-product state is given by}
\begin{equation}
|\psi_{\mathrm{coh}}\rangle=|\alpha\rangle\otimes|\downarrow\rangle^{\otimes N},
\label{eq:coherent_state}
\end{equation}
\Chen{where $|\alpha\rangle =
e^{-|\alpha|^2/2}
 \sum_{n=0}^{\infty}
\frac{\alpha^n}{\sqrt{n!}}|n\rangle$ is a coherent state of the cavity mode with amplitude $\alpha$}. This state explicitly breaks the $\mathbb{Z}_2$ symmetry, since
\begin{equation}
\hat{P}|\psi_{\mathrm{coh}}\rangle
=
|-\alpha\rangle\otimes|\downarrow\rangle^{\otimes N}.
\end{equation}
Its degree of parity breaking can be characterized by
\begin{equation}
|\langle\psi_{\mathrm{coh}}|\hat{P}|\psi_{\mathrm{coh}}\rangle|
=
|\langle\alpha|-\alpha\rangle|
=
e^{-2|\alpha|^{2}},
\label{eq:parity_offset_cs}
\end{equation}
which decreases monotonically with $|\alpha|$. A smaller parity expectation value corresponds to stronger symmetry breaking. 

The second class is a family of \textcolor{black}{cat-product states}, which \CH{has been} experimentally realized in microwave-cavity and circuit-QED platforms~\cite{Deleglise2008Nature,Vlastakis2013Science,Wang2016Science}:
\begin{equation}
|\psi_{\mathrm{cat}}\rangle
=
\textcolor{black}{\mathcal{N}\left(|\alpha\rangle+e^{i\theta}|-\alpha\rangle\right)\otimes|\downarrow\rangle^{\otimes N}},
\label{eq:cat_state}
\end{equation}
where 
\begin{equation}
\mathcal{N}
=
\textcolor{black}{\frac{1}{\sqrt{2\left(1+\cos\theta\,e^{-2|\alpha|^2}\right)}}}
\end{equation}
is the normalization factor. This state continuously interpolates between the coherent composition of the parity-symmetric and parity-antisymmetric cat components. The corresponding parity expectation value is
\begin{equation}
|\langle\psi_{\mathrm{cat}}|\hat{P}|\psi_{\mathrm{cat}}\rangle|
=
\left|
\textcolor{black}{\frac{e^{-2|\alpha|^{2}}+\cos\theta}
{1+\cos\theta\,e^{-2|\alpha|^{2}}}}
\right|.
\label{eq:parity_offset_ss}
\end{equation}
Evidently, the \textcolor{black}{cat-product family} is controlled by two parameters: the coherent amplitude $\alpha$, which controls the magnitude of the classical displacement, and the phase $\theta$, which controls the interference between the two parity-related coherent components. \textcolor{black}{At $\theta=0$ and $\theta=\pi$, the cat-product state becomes a parity eigenstate: the even cat for $\theta=0$ and the odd cat for $\theta=\pi$. Accordingly, $\langle \hat P\rangle=+1$ and $-1$, respectively.}
\CH{In these limits, the initial state is effectively confined to a definite parity sector, without any parity-breaking dependence on $\alpha$.} 
\textcolor{black}{Furthermore, Eq.~\ref{eq:parity_offset_ss} shows that for $0<\theta/(2\pi)<1/4$, increasing $\alpha$ monotonically decreases the magnitude of the parity expectation value, corresponding to a monotonic enhancement of the initial parity asymmetry. We therefore focus on}
\begin{equation}
\textcolor{black}{0<\frac{\theta}{2\pi}<\frac{1}{4}},
\end{equation}
\Chen{where varying $\alpha$ provides a systematic way to control the initial degree of parity breaking and compare the subsequent relaxation dynamics.}

\begin{figure}
\includegraphics[width=80mm]{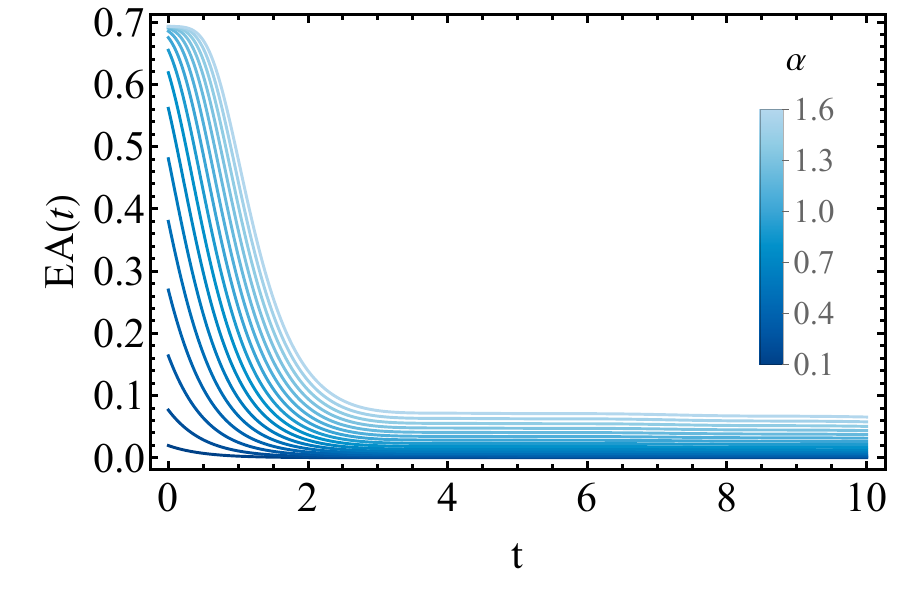}
\caption{\Chen{Absence of the quantum Mpemba effect (QME) for coherent-product states in the normal phase. The curves show the time evolution of the EA for the coherent-product initial states defined in Eq.~(\ref{eq:coherent_state}), with $\alpha=0.1\text{-}1.6$. The values of $\alpha$ increase from bottom to top at $t=0$. All trajectories relax monotonically toward the steady state without crossing, indicating the absence of the QME for this family of initial states. The parameters are $\omega=\omega_0=1$, $\kappa=2$, and $\lambda=0.1$, corresponding to the normal phase.}}
\label{Fig:1}
\end{figure}

\begin{figure*}[!t]
\includegraphics[width=110mm]{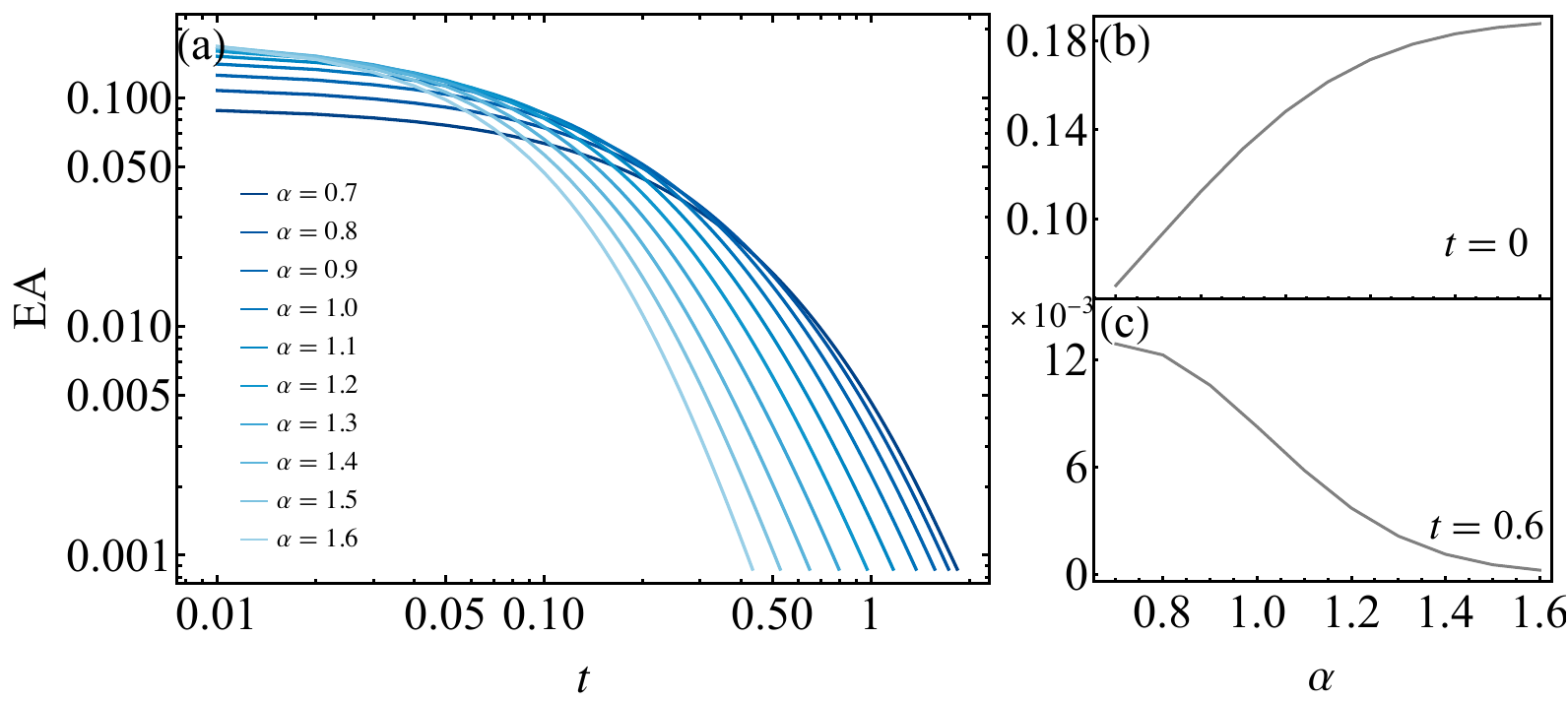}
\caption{\Chen{Quantum Mpemba effect induced by cat-product initial states. The initial state is $|\psi_{\rm cat}\rangle
=\mathcal{N}(|\alpha\rangle+\textcolor{black}{e^{i\theta}}|-\alpha\rangle)
\otimes|J,-J\rangle$, where $\mathcal{N}$ is the normalization factor.
(a) Time evolution of the entanglement asymmetry (EA) for different $\alpha$ in the normal phase ($\lambda=0.1$), shown on a log-log scale.
Although states with larger $\alpha$ have larger initial EA, their relaxation trajectories cross those with smaller $\alpha$, demonstrating an inversion of the relaxation hierarchy and hence the QME. (b) Initial EA as a function of $\alpha$. (c) EA at $t=0.6$ as a function of $\alpha$.
Panels (b) and (c) focus on $0.7\leq\alpha\leq1.6$,
where the inversion of the relaxation hierarchy is
clearly visible. Crossings are also observed for certain values of $\alpha$ below $0.7$, although they are less pronounced than those in the range shown. The parameters are \textcolor{black}{$\theta/(2\pi)=0.1$}, $\omega=\omega_0=1$, $\kappa=2$, and $\lambda=0.1$.
The photon cutoff is $n_p=20$, and the atom number is $N=10$.}
}
\label{Fig:2}
\end{figure*}

\subsection{Initial-state dependence}

We first consider the normal phase \CH{$\lambda < \lambda_c$}, where the steady state is parity-symmetric and unique. Figure~\ref{Fig:1} shows the time evolution of $\mathrm{EA}(t)$ for coherent-product states with different amplitudes $\alpha$. \CH{At all times $t$, }the EA increases monotonically with $\alpha$ -- all trajectories relax smoothly toward the steady state without crossing. The relaxation ordering is therefore preserved, and no QME is observed for this class of initial states.

The absence of crossings reflects the simple relaxation structure of coherent-product states under photon loss. In such states, parity breaking is mainly associated with a single cavity displacement, \CH{with the coherent states being eigenvectors} $a|\alpha\rangle=\alpha|\alpha\rangle$. Photon loss therefore damps this displacement in a regular manner. Varying $\alpha$ changes the initial magnitude of the EA, but within this coherent-product family it does not introduce distinct interference structures or substantially different projections onto the slow relaxation modes. The EA trajectories therefore remain ordered throughout the evolution, and no inversion of relaxation ordering occurs.

By contrast, qualitatively different behavior appears for the \textcolor{black}{cat-product states} defined in Eq.~(\ref{eq:cat_state}). As shown in Fig.~\ref{Fig:2}(a), the trajectories of $\mathrm{EA}(t)$ exhibit prominent crossings over a finite range of $\alpha$ \CH{at early times}, indicating an inversion of the relaxation ordering. States with larger initial EA can evolve to smaller EA at later times, which constitutes the signature of QME.
{\color{black}
To elucidate the microscopic origin of this anomalous relaxation behavior for cat-product states, we analyze the Liouvillian spectral decomposition. We label the stationary mode by $k=1$, with $\mu_1=0$ and \textcolor{black}{$R_1=\rho_{\mathrm{steady}}/\|\rho_{\mathrm{steady}}\|_{\rm HS}$}, and order the nonstationary modes by increasing decay rate, $0<-\mathrm{Re}\,\mu_2\leq-\mathrm{Re}\,\mu_3\leq\cdots.$ The density matrix can be expanded as 
\begin{equation}
\rho(t)=\rho_{\rm steady}+\sum_{k\ge2} c_k\, e^{\mu_k t} R_k,
\label{eq:liouvillian_expansion}
\end{equation}
where $\mu_k$ are nonzero Liouvillian eigenvalues and $R_k$ ($L_k$) are the corresponding right (left) Liouvillian eigenoperators, as defined in Appendix B. \textcolor{black}{We fix their scaling by imposing $\|R_k\|_{\rm HS}=1$ together with the biorthogonality condition $\mathrm{Tr}(L_k^\dagger R_j)=\delta_{kj}$.} The dependence on the initial state enters through the overlap coefficients,
\begin{equation}
c_k=\mathrm{Tr}\!\left(L_k^\dagger\rho_0\right).
\label{eq:ck_def}
\end{equation}
where $\rho_0 =\rho(t = 0)$. The slowest relaxation channel is controlled by the mode with the smallest nonzero decay rate $-\Re\,\mu_k$. 
}

\begin{figure}
\includegraphics[width=0.48
\textwidth]{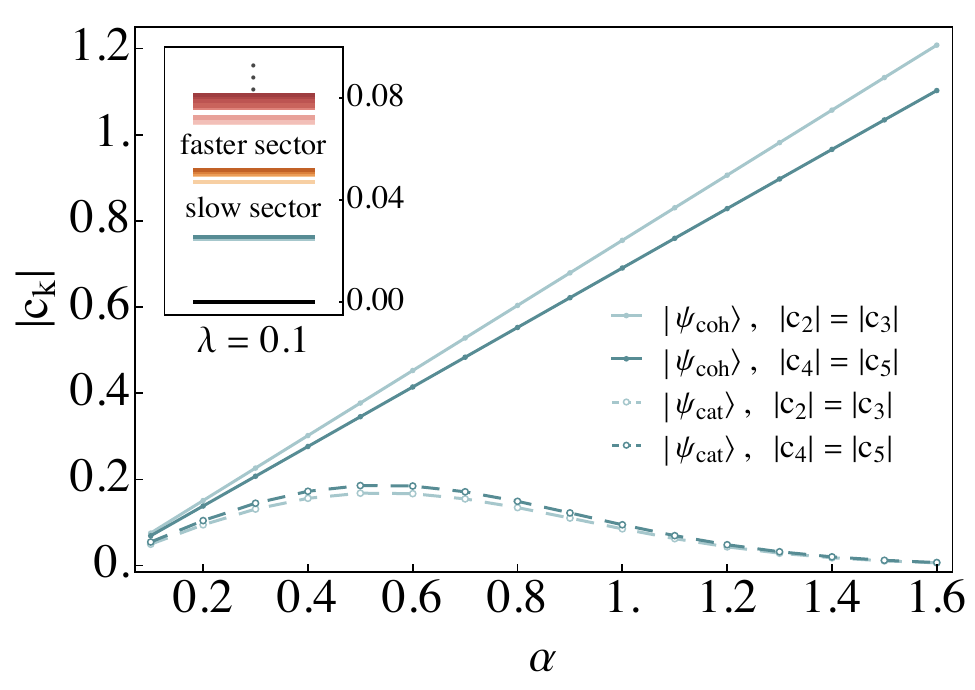}
\caption{Initial-state-dependent filtering of the leading slow Liouvillian modes. Overlaps $|c_k|$ of the coherent-product and cat-product initial states with the leading slow modes as a function of $\alpha$. For the coherent-product states, $|c_2|=|c_3|$ and $|c_4|=|c_5|$ increase approximately linearly with $\alpha$, whereas for the cat-product states both pairs are suppressed at larger $\alpha$. The collective suppression of the leading slow-mode overlaps reduces the contribution of the slow relaxation channels, shifting the dynamics toward faster-decaying modes and thereby providing a spectral explanation for the QME. The inset shows the Liouvillian decay-rate spectrum $-\mathrm{Re}\,\mu_k$ at $\lambda=0.1$, showing the separation between the leading slow sector ($k=2,\ldots,5$) and the faster decaying modes. The spacing between the two slow-mode pairs in the inset is enlarged for visual clarity, while the degeneracy within each pair is preserved. The parameters are the same as those in Fig.~\ref{Fig:2}.}
\label{Fig:3}
\end{figure}

As shown in Fig.~\ref{Fig:3}, \Chen{the overlaps with the leading slow Liouvillian modes depend strongly on the initial-state structure. For the coherent-product states, both $|c_2|=|c_3|$ and $|c_4|=|c_5|$ increase approximately linearly with $\alpha$. In contrast, for the cat-product states, these overlaps first increase and then become suppressed as $\alpha$ increases. Importantly, this suppression is not restricted to a single mode, but occurs collectively for the leading slow modes $k=2,\ldots,5$. 
Consequently, the relaxation of the cat-product states becomes less dominated by the slow sector and is instead governed by faster-decaying Liouvillian modes. In the parameter regime where the EA trajectories exhibit the Mpemba crossing, this filtering of the leading slow 
relaxation channels provides a spectral explanation for the accelerated relaxation of the cat-product states and the resulting QME.}

{\color{black}
The markedly different relaxation behavior of the coherent- and cat-product initial states can be understood from their overlaps with the leading slow Liouvillian modes. In the normal phase, the low-excitation condition ($(\langle J_z\rangle+N/2)/N\ll1$) allows us to linearize the dynamics around the polarized state (Appendix~\ref{app:linear_fluctuation}). The resulting first-order fluctuation modes are linear combinations of cavity and spin operators and provide an approximate description of the leading slow relaxation. For both initial-state families, the atoms are prepared in $|J,-J\rangle$. Taking the matrix element in this atomic state removes the linear spin terms, leaving the cavity terms that determine the initial overlaps. For a representative slow left eigenmode $L_2$, this gives
\begin{equation}
\ell_2
\equiv
\langle J,-J|L_2|J,-J\rangle_{\rm spin}
\simeq u_2 a+v_2 a^\dagger .
\label{eq:l2_eff}
\end{equation}
Since $c_2=\mathrm{Tr}(L_2^\dagger\rho_0)$, the overlap probes $\ell_2^\dagger$. For real $\alpha$, using 
$\langle a\rangle=\langle a^\dagger\rangle=\alpha$, this immediately gives
\begin{equation}
|c_2^{\rm coh}(\alpha)|\sim C_{\rm coh}\alpha ,
\label{eq:c2_coherent}
\end{equation}
in agreement with the numerical results in Fig.~\ref{Fig:3}. The same effective linear profile also accounts for the qualitatively 
different behavior of the cat-product states. Numerically, $L_2$ is 
parity odd, $P L_2 P^\dagger=-L_2$. The diagonal contributions from the two parity-related branches $|\alpha\rangle$ and $|-\alpha\rangle$ therefore cancel, leaving only the inter-branch coherence. For real $\alpha$, the linear profile in Eq.~(\ref{eq:l2_eff}) gives
\begin{equation}
\langle\alpha|\ell_2^\dagger|-\alpha\rangle
\simeq(u_2^*-v_2^*)\alpha e^{-2\alpha^2},
\label{eq:cross_element}
\end{equation}
where $\langle\alpha|-\alpha\rangle=e^{-2\alpha^2}$. Combining the two 
inter-branch contributions with the relative phase and normalization of the \textcolor{black}{cat-product state} then gives
\begin{equation}
\left|\textcolor{black}{c_{2}^{\rm cat}}(\alpha,\theta)\right|
\sim \textcolor{black}{C_{\rm cat}}
\frac{|\sin\theta|\,\alpha e^{-2\alpha^{2}}}{
1+\cos\theta\,e^{-2\alpha^{2}}},
\label{eq:c2_cat}
\end{equation}
where $\textcolor{black}{C_{\rm cat}}$ is independent of $\alpha$ and $\theta$. Equations~(\ref{eq:c2_coherent}) and (\ref{eq:c2_cat}) make the physical origin of the mode-filtering mechanism particularly transparent. The coherent-product overlap grows approximately linearly with $\alpha$, whereas for cat-product states the exponential factor $e^{-2\alpha^2}$ eventually outweighs the linear growth, suppressing the slow-mode overlap at larger $\alpha$.  Meanwhile, the factor $|\sin\theta|$ implies that the overlap with the parity-odd slow mode vanishes at $\theta=0$ and $\theta=\pi$, corresponding to the even- and odd-cat states, respectively. In both cases, the associated density matrix is parity even and therefore has no projection onto the parity-odd Liouvillian sector. Together with the simultaneous suppression of the other leading slow modes shown in Fig.~\ref{Fig:3}, this selectively reduces the weight of the slow relaxation sector and shifts the relaxation dynamics toward faster-decaying Liouvillian modes, thereby giving rise to the Mpemba inversion.}



\begin{figure}
\includegraphics[width=0.4\textwidth]{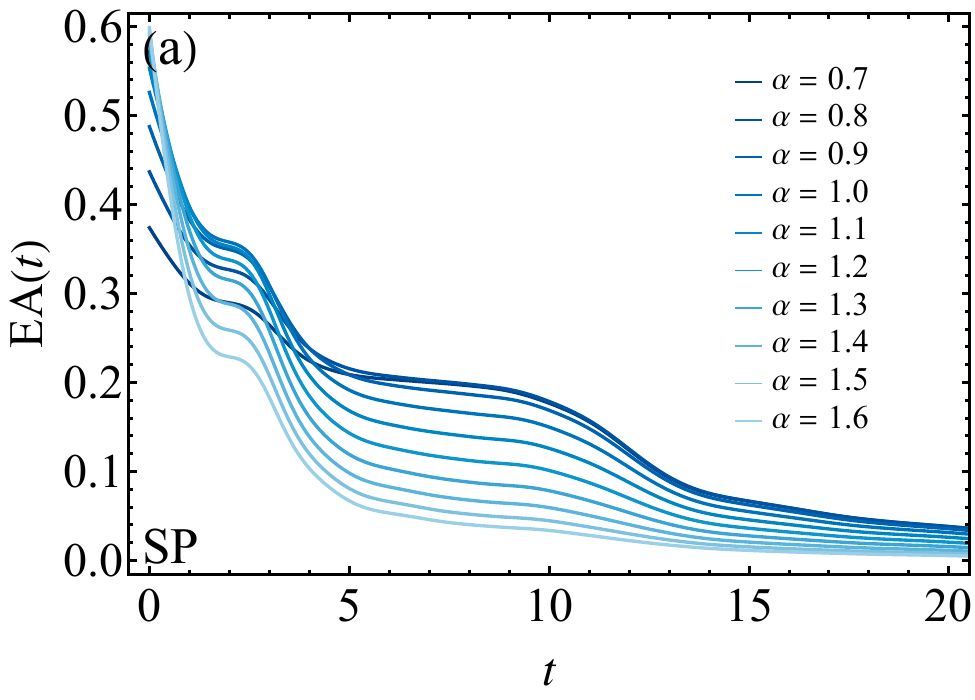}
\hspace{0.8cm}
\includegraphics[width=0.4\textwidth]{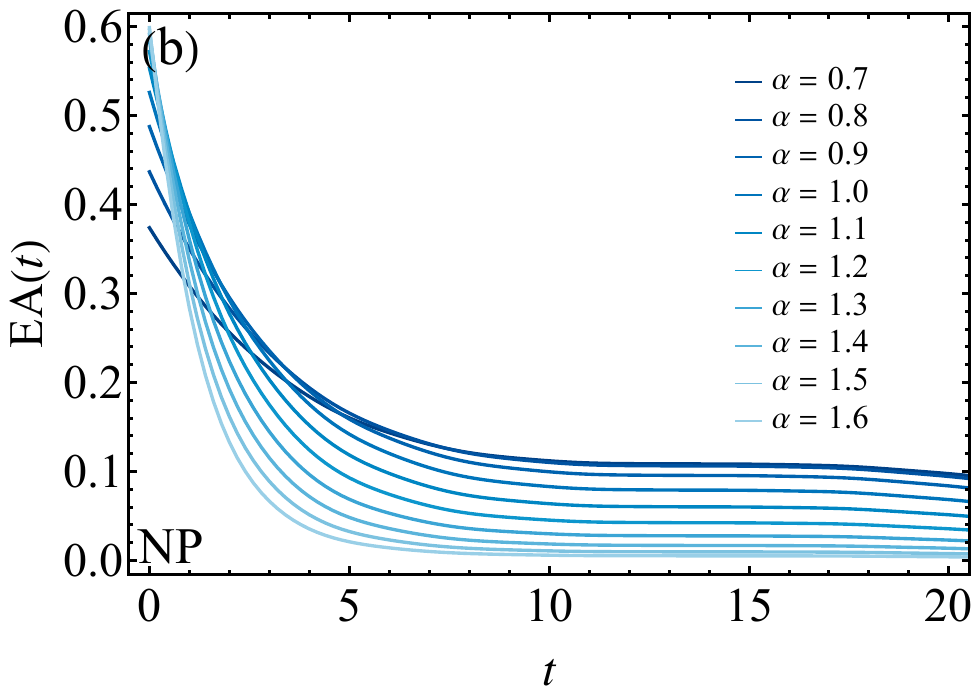}
\caption{
\CH{Relaxation profiles in the superradiant vs. normal phases.} (a) Superradiant regime with $\lambda=0.65>\lambda_c$. (b) Normal regime with $\lambda=0.1<\lambda_c$. Each curve corresponds to a \textcolor{black}{cat-product initial state} with a different amplitude $\alpha$, as indicated in the legends. The other parameters are \textcolor{black}{$\theta/(2\pi)=0.2$}, $\omega=\omega_0=1$, and $\kappa=0.1$. The photon cutoff is $n_{\mathrm{p}}=20$, and the atom number is $N=10$.}
\label{Fig:4}
\end{figure}

\Chen{The emergence of the Mpemba crossings can be understood from the increasing distinguishability of the two parity-related coherent components. Their overlap is $|\langle\alpha|-\alpha\rangle|=e^{-2|\alpha|^2}.$ For small $\alpha$, the two branches strongly overlap, and the coherence-induced filtering of the slow Liouvillian sector remains weak. As $\alpha$ increases, the branch overlap decreases exponentially. Consequently, the interference between the two parity-related components increasingly suppresses their projections onto the slow modes, consistent with the $\alpha$-dependence of the slow-mode coefficients in Eq.~\ref{eq:c2_cat}. The resulting reduction of the slow-sector contribution enhances the relative importance of the faster-decaying modes in the relaxation, thereby facilitating the emergence of Mpemba crossings. Importantly, the factor $e^{-2|\alpha|^2}$ does not define a sharp onset of the Mpemba effect. Indeed, \textcolor{black}{for the $\kappa=2$ parameters of Fig.~\ref{Fig:2},} a Mpemba crossing is already found between the trajectories for $\alpha=0.55$ and $0.65$, below the range where the crossings become visually prominent in Fig.~\ref{Fig:2}. A Mpemba crossing is a relative dynamical property of two initial states and therefore cannot be determined solely from the suppression of the slow-mode overlaps.}

\subsection{Low-lying Liouvillian structure and multi-timescale relaxation}

\begin{figure*}[!t]
\includegraphics[width=120mm]{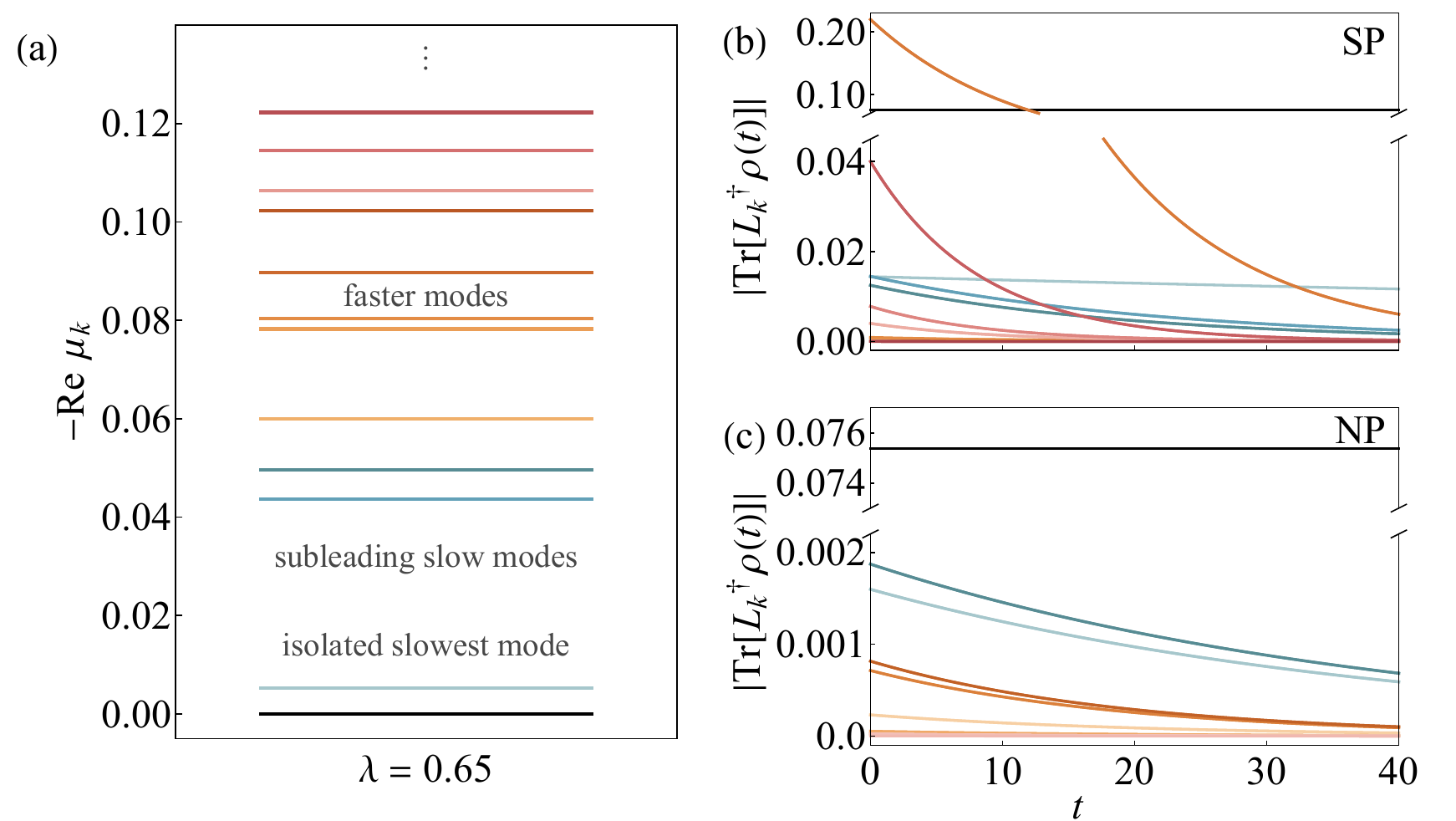}
\caption{\textcolor{black}{Comparison between the normal and superradiant regimes. (a) Low-lying Liouvillian decay rates $-\mathrm{Re}\,\mu_k$ for the superradiant regime at $\lambda=0.65>\lambda_c$, showing an isolated slowest nonstationary mode, a group of subleading slow modes, and faster modes. (b,c) Time-dependent modal amplitudes $|\mathrm{Tr}[L_k^\dagger\rho(t)]|=|c_k|e^{\mathrm{Re}\,\mu_k t}$ of the first $20$ Liouvillian modes in the superradiant ($\lambda=0.65$) and normal ($\lambda=0.1$) regimes, respectively, using Hilbert--Schmidt-normalized right eigenoperators, $\|R_k\|_{\rm HS}=1$. The stationary mode $\mu_1=0$ gives the time-independent contribution. A broken vertical axis is used in (b,c) to display the stationary and nonstationary contributions on the same scale. Here $\alpha=0.8$, and the other parameters are the same as those in Fig.~\ref{Fig:4}.}}
\label{Fig:5}
\end{figure*}

We next examine how the relaxation profile is affected by the structure of the low-lying Liouvillian modes. To this end, we compare two representative coupling strengths on the two sides of the dissipative Dicke transition: \CH{the normal and the superradiant phases}. For the finite systems considered here, the steady state remains unique and parity symmetric, but the low-lying Liouvillian spectrum can change substantially with the system parameters. A clearer separation of slow-mode decay rates and mode weights can then give rise to a more structured multi-timescale relaxation.

As shown in Fig.~\ref{Fig:4}(a), the relaxation dynamics in the superradiant regime become more \CH{nuanced than in the normal regime shown in Fig.~\ref{Fig:4}(b). The superradiant EA trajectories display more pronounced separations over an extended time window, and the relaxation is no longer characterized by a nearly single-stage decay.} As such, the dynamics reflect the competition among several slow Liouvillian modes.
\Chen{Figure~\ref{Fig:5} provides a mode-resolved view of this difference. For the representative superradiant point $\lambda=0.65$, Fig.~\ref{Fig:5}(a) reveals a hierarchical structure in the low-lying Liouvillian spectrum, with an isolated slowest nonstationary mode, several subleading slow modes, and a group of faster modes. As shown in Fig.~\ref{Fig:5}(b), several of these modes remain dynamically relevant during the relaxation, and their time-dependent contributions cross as the relative importance of different modes changes. Such mode competition and switching of the dominant contributions provide a modal origin of the multiple relaxation timescales observed in the EA dynamics. In contrast, at the representative normal-phase point [Fig.~\ref{Fig:5}(c)], the nonstationary modal contributions are less pronounced and show a much weaker competition among different modes, \textcolor{black}{consistent with the nearly single-stage EA relaxation in Fig.~\ref{Fig:4}(b).} Together with the initial-state dependence of the mode coefficients discussed above, these results show that the relaxation dynamics is determined jointly by the initial modal weights and the decay rates of the low-lying Liouvillian modes.}

\Chen{Taken together, the results of this section establish a mode-resolved picture of the quantum Mpemba effect in the dissipative Dicke model. According to the Liouvillian decomposition in Eq.~\ref{eq:liouvillian_expansion}, the time-dependent modal magnitude satisfies
$\left|\mathrm{Tr}\!\left[L_k^\dagger\rho(t)\right]\right|
=|c_k|e^{\mathrm{Re}\,\mu_k t},
$ showing explicitly how the initial modal coefficient $c_k$ and the corresponding decay rate $-\mathrm{Re}\,\mu_k$ jointly determine each modal contribution.
The initial-state dependence of $c_k$ shown in Fig.~\ref{Fig:3} demonstrates that the \textcolor{black}{cat-product state} can suppress its coupling to the dominant slow relaxation channels, thereby modifying the relative contributions of the low-lying Liouvillian modes. This redistribution of modal weights provides the spectral mechanism underlying the reversal of the relaxation ordering. The detailed temporal profile, however, depends on the structure of the dynamically relevant modes. \textcolor{black}{For the representative parameters considered here, the weaker mode competition in the normal regime is associated with the nearly single-stage EA decay in Fig.~\ref{Fig:4}(b), whereas the superradiant regime exhibits a more structured hierarchy of dynamically relevant Liouvillian modes [Fig.~\ref{Fig:5}(b)] and is associated with the more pronounced multi-timescale relaxation profile in Fig.~\ref{Fig:4}(a).}}

\Chen{The dependence of this relaxation structure on the dissipative dynamics is further illustrated in Appendix~\ref{Appendix:A}. When collective atomic decay through the $J_-$ channel is included, the multi-timescale features of the EA dynamics are strongly weakened compared with the case of cavity loss alone. The mode-filtering mechanism based on the initial modal weights and Liouvillian decay rates therefore provides the general spectral framework, while the detailed distribution and competition of the relevant modes depend on the system parameters and dissipation channels.}

\section{Parameter-space structure of the quantum Mpemba effect}
\label{sec:PhaseDiagram}

\begin{figure*}
\includegraphics[width=150mm]{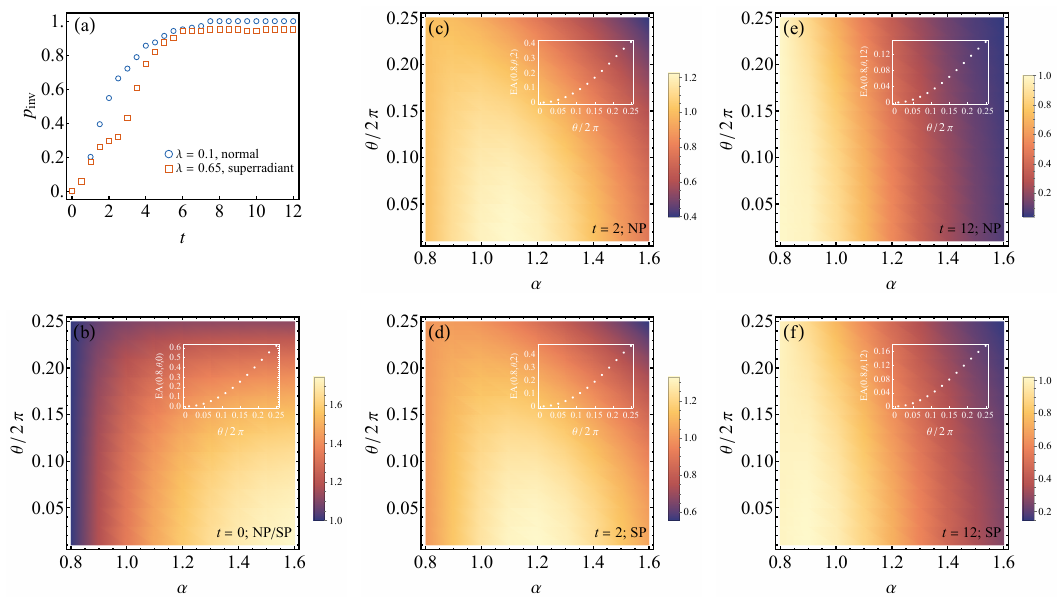}
\caption{\textcolor{black}{Robustness of the QME beyond isolated fine-tuned initial states. (a) Fraction $p_{\mathrm{inv}}$ of sampled $(\alpha,\theta/2\pi)$ points exhibiting an inverted relaxation ordering relative to the $\alpha=0.8$ reference, shown as a function of time for the normal (NP, $\lambda=0.1$) and superradiant (SP, $\lambda=0.65$) regimes. A point is counted as inverted when $\mathrm{EA}(\alpha,\theta,t)<\mathrm{EA}(0.8,\theta,t)$ for $\alpha>0.8$. (b) Initial normalized EA at $t=0$, common to the NP and SP cases. (c,d) Normalized EA at the intermediate time $t=2$ in the NP and SP regimes, respectively. (e,f) Corresponding results at the later time $t=12$. In panels (b)--(f), the EA is normalized by $\mathrm{EA}(0.8,\theta,t)$ at the corresponding time, and the insets show the unnormalized reference value $\mathrm{EA}(0.8,\theta,t)$ as a function of $\theta/2\pi$. The initial ordering with increasing $\alpha$ is progressively reversed during the evolution over an extended region of the $(\alpha,\theta/2\pi)$ plane, with distinct transient behavior in the two regimes. Other parameters are $\omega=\omega_0=1$, $\kappa=0.1$, and $0<\theta/(2\pi)<1/4$.}}
\label{Fig:6}
\end{figure*}

\CH{We now show that the QME in our open Dicke model is not
restricted to specially selected initial states or time
snapshots. This will be evidenced in relaxation-order
inversion over a broad region} of the
\textcolor{black}{cat-product parameter space}.
We therefore examine the \textcolor{black}{cat-product states}
introduced in Eq.~(\ref{eq:cat_state})  across the explored
\textcolor{black}{$(\alpha,\theta)$} parameter region.
The coherent amplitude $\alpha$ controls the displacement
of the two coherent components, whereas $\theta$ sets their
relative phase and thereby modifies the internal structure
of the initial superposition. This two-parameter analysis
allows us to determine whether the relaxation inversion
is confined to isolated initial states and to separate
the roles of the coherent amplitude and relative phase
in shaping its parameter-space structure.

{\color{black}
To identify relaxation-order inversion, we first examine
the initial EA as a function of $\alpha$ and $\theta$.
For the real amplitudes considered here, the initial EA
of the pure cat-product states is exactly
\begin{align}
\mathrm{EA}(0)
&=
\ln\left[
\frac{2}{1+p_0^2}
\right],
\\
p_0
&\equiv
\langle\psi_{\rm cat}|\hat P|\psi_{\rm cat}\rangle
=
\frac{e^{-2\alpha^2}+\cos\theta}
{1+\cos\theta\,e^{-2\alpha^2}}.
\label{eq:initial_EA_cat}
\end{align}
Introducing $x=e^{-2\alpha^2}$ and $c=\cos\theta$, one obtains
\begin{equation}
p_0=\frac{x+c}{1+cx},
\qquad
\frac{\partial p_0}{\partial x}
=
\frac{1-c^2}{(1+cx)^2}.
\label{eq:parity_monotonicity}
\end{equation}
Within the interval studied here, $0<\theta<\pi/2$,we have $0<c<1$ and $p_0>0$. Since $x$ decreases with $\alpha$, $p_0$ decreases while the initial EA increases monotonically with $\alpha$ at fixed $\theta$. At fixed $\alpha>0$, the initial EA also increases monotonically with $\theta$ within this interval.

The initial EA alone does not determine the relaxation order, which also depends on the weights of the slowly decaying Liouvillian modes excited by the initial state. The different influence of the two parameters on the
subsequent relaxation follows from Eq.~\ref{eq:c2_cat}.
The leading slow-mode coefficient contains the exponential
factor $e^{-2\alpha^2}$, whereas the relative phase enters
through trigonometric interference factors. Within the
approximation underlying that expression, increasing
$\alpha$ over the reference-state comparison range
$\alpha\geq0.8$ suppresses the magnitude of the leading
parity-odd slow-mode overlap. By contrast, increasing
$\theta$ at fixed $\alpha$ increases the magnitude of
this overlap within the phase interval studied here.
Thus, increasing $\alpha$ combines a larger initial EA
with a smaller slow-mode overlap, whereas increasing
$\theta$ increases both. 

Figure~\ref{Fig:6}  characterizes the development of relaxation-order inversion across the cat-product parameter space. Panel (a) shows the fraction of sampled initial states
exhibiting an inverted EA ordering relative to the
reference state with $\alpha=0.8$.
For each sampled $\theta$, we compare the sampled initial
states with $\alpha>0.8$ against the reference state with
$\alpha=0.8$ at the same $\theta$.
We denote by $N_{\rm tot}$ the total number of these sampled
states across all sampled values of $\theta$.
This count excludes the reference states and is independent
of time.
At time $t$, $N_{\rm inv}(t)$ counts the states within this
same set whose EA is smaller than that of the corresponding
reference state at that time.
The instantaneous inversion fraction shown in panel (a)
is then
\begin{equation}
p_{\rm inv}(t)=\frac{N_{\rm inv}(t)}{N_{\rm tot}}.
\end{equation}
The monotonic initial ordering ensures $p_{\rm inv}(0)=0$. Panel~(b) presents the initial ordering, which is common to both coupling regimes. In each color map, the EA is normalized by its value at $\alpha=0.8$ for the same $\theta$ and time. For $\alpha>0.8$, the normalized EA
is initially greater than unity; values below unity
at later times therefore indicate an inversion relative
to the corresponding reference state. The intermediate-time maps at $t=2$ (Figs.~\ref{Fig:6}(c,d)) show the developing redistribution of the normalized EA. At $t=12$ (Figs.~\ref{Fig:6}(e,f)), the initial ordering relative to the reference states is reversed over an extended region of parameter space: states with larger initial EA can develop smaller EA than their respective reference states, corresponding to faster parity restoration according to the EA diagnostic. The time dependence of $p_{\rm inv}$ and the evolution of the parameter-space maps demonstrate that the Mpemba inversion develops over a finite region of the initial-state manifold.

The inversion is present in both the normal and superradiant
regimes, but its distribution in the $(\alpha,\theta)$
plane differs between the two regimes at both intermediate
and late times (Figs.~\ref{Fig:6}(c,d) and \ref{Fig:6}(e,f), respectively). This difference is consistent with the Liouvillian spectral analysis discussed above. Since the low-lying relaxation modes have different
hierarchical structures in the two regimes, the balance
between the suppressed slow-mode contribution and the
subleading-mode contributions is reflected differently
across the initial-state parameter space.

\section{Conclusion}
\label{sec:conclusion}

In this work, we investigated symmetry-restoration quantum Mpemba dynamics in the weakly $\mathbb{Z}_2$-symmetric dissipative Dicke model. Using the EA and Liouvillian spectral decomposition, we showed that anomalous parity-restoration ordering can arise even without strong symmetry conservation or an extensive symmetry-sector structure. While coherent-product states exhibit conventional relaxation with no EA crossings, \textcolor{black}{cat-product states} display clear symmetry-restoration Mpemba behavior, where states with larger initial parity asymmetry restore the weak $\mathbb{Z}_2$ symmetry more rapidly at later times.

The spectral analysis identifies the underlying mechanism as initial-state mode filtering. The coherent superposition of two parity-related coherent states suppresses the projection onto the leading slow Liouvillian modes, allowing faster subleading modes to contribute more prominently to the observable parity-restoration dynamics. The same filtering mechanism operates across the studied parameter regimes, while the low-lying Liouvillian structure controls whether the EA dynamics remain nearly single-stage or develop a multi-timescale profile. Mapping the \textcolor{black}{$(\alpha,\theta/2\pi)$} plane further shows that the effect extends over a finite region of the \textcolor{black}{cat-product manifold}, rather than being restricted to isolated fine-tuned points. Additional collective atomic decay smooths the multi-timescale profile but does not eliminate the Mpemba-type crossings, indicating that the crossing mechanism is more robust than the detailed relaxation shape.

From an experimental perspective, testing the present symmetry-restoration diagnostic requires estimating the two second moments $\mathrm{Tr}[\rho^2(t)]$ and $\mathrm{Tr}[\rho_P^2(t)]$, rather than performing full state tomography. For finite systems, these quantities can in principle be accessed through randomized-measurement and classical-shadow protocols. Related techniques have already been used to measure entanglement asymmetry and observe the QME in a trapped-ion quantum simulator~\cite{Joshi2024ObservingQME}. Cavity- and circuit-QED platforms provide natural settings for preparing photonic cat states and realizing dissipative Dicke dynamics~\cite{Deleglise2008Nature,Vlastakis2013Science,Wang2016Science,Felicetti2018PRA}, while programmable quantum processors offer an alternative route for implementing controlled open-system dynamics~\cite{chen2023high,koh2025interacting,chen2026robust,xu2026fractional}. Extending these measurement protocols to the full dissipative many-body density matrix remains an important experimental challenge. Overall, these results establish initial-state mode filtering as a microscopic mechanism for quantum Mpemba relaxation in weakly symmetric open systems, and demonstrate its realization in the paradigmatic light--matter setting of the dissipative Dicke model.

\bibliography{refs}

\newpage
\appendix
\section{Effect of collective atomic decay}
\label{Appendix:A}

In the main text, we first analyzed the dissipative dynamics with cavity loss. Here, we examine how the relaxation dynamics change when an additional collective atomic decay channel is included. The master equation becomes
\begin{equation}
\frac{d\rho}{dt}=-i[H,\rho]+\mathcal D[L_\kappa]\rho+\mathcal D[L_\gamma]\rho ,
\end{equation}
where $L_\kappa=\sqrt{\kappa} a, L_\gamma=\sqrt{\gamma} J_-$.
The collective decay channel preserves the permutation-symmetric Dicke subspace and also respects the same weak $\mathbb{Z}_2$ symmetry: $J_-$ changes sign under the parity transformation and therefore connects Hilbert-space parity sectors, but the Lindblad dissipator is unchanged because $\mathcal D[-J_-]=\mathcal D[J_-]$.
\Chen{Figure~\ref{Fig:S1} shows the EA dynamics in the presence of both cavity loss and collective atomic decay. In the normal regime, the relaxation dynamics is only weakly affected by the additional $J_-$ channel, and the Mpemba-type crossings remain visible. In contrast, collective atomic decay has a more pronounced effect in the superradiant regime, where the multistage relaxation observed under cavity loss alone is substantially smoothed out. We have also checked the dependence on the collective decay rate for $\gamma$=0.05, 0.1, and 0.2. Within the range considered, increasing $\gamma$ does not lead to qualitatively new behavior: the normal-regime dynamics remains only weakly affected, while the multistage relaxation in the superradiant regime is progressively suppressed. The Mpemba-type crossings persist throughout the range considered.}

\begin{figure}[ht!]
\includegraphics[width=0.4\textwidth]{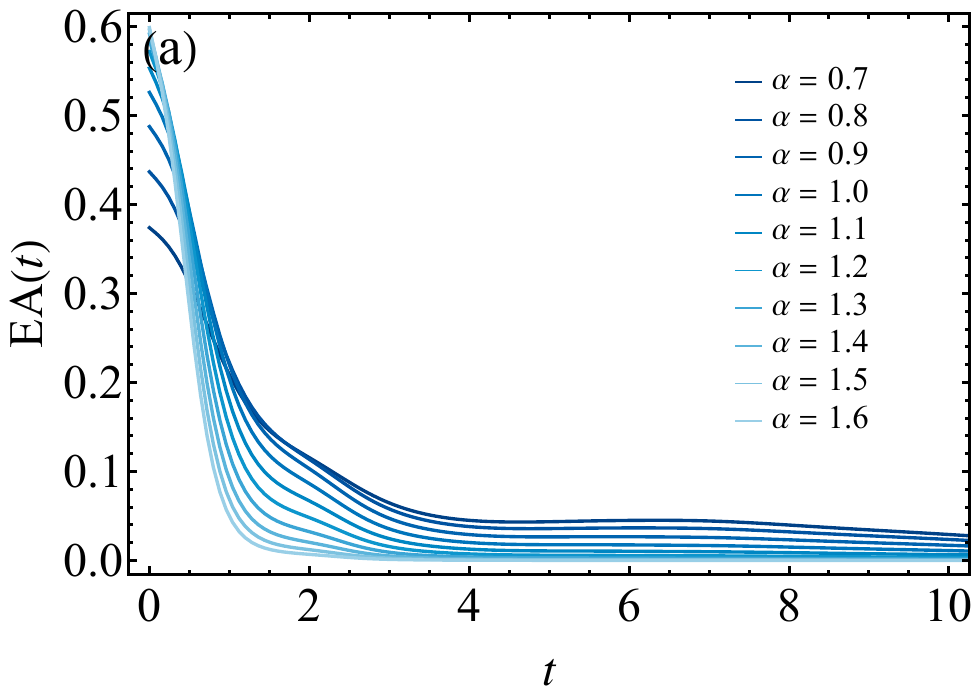}
\hspace{0.8cm}
\includegraphics[width=0.4\textwidth]{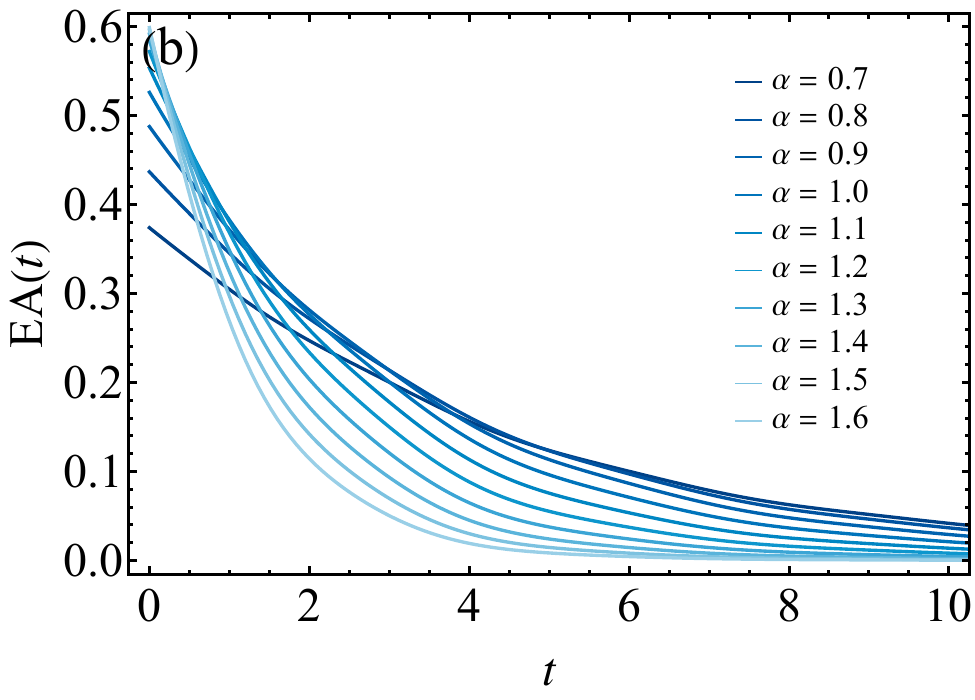}
\caption{\Chen{Collective atomic decay erases the multistage profile but leaves the Mpemba crossings intact. EA dynamics in the presence of both cavity loss and collective atomic decay for (a) the superradiant regime ($\lambda=0.65$) and (b) the normal regime ($\lambda=0.1$). Each curve corresponds to a \textcolor{black}{cat-product initial state} with different coherent amplitude $\alpha$. 
Compared with the result with cavity loss in Fig.~\ref{Fig:4}, the inclusion of collective atomic decay smooths out the step-like, multi-stage relaxation in the superradiant regime, while the Mpemba-type crossings remain visible.
The other parameters are \textcolor{black}{$\theta/(2\pi)=0.2$}, $\omega=\omega_0=1$, and $\gamma=\kappa=0.1$.}}
\label{Fig:S1}
\end{figure}

\Chen{This different sensitivity to collective atomic decay can be understood from the atomic excitation in the two regimes. In the normal regime, the collective spin remains close to $|J,-J\rangle$, for which $J_-|J,-J\rangle=0$, so that the additional atomic decay channel has only a weak effect. In the superradiant regime, by contrast, the atomic degree of freedom is appreciably excited and participates more strongly in the relaxation dynamics. Under cavity loss alone, dissipation acts directly on the photonic mode, while the atomic component relaxes indirectly through its coupling to the cavity field. The $J_-$ channel provides an additional direct relaxation pathway for the atomic component, thereby reducing the separation between relaxation timescales and smoothing the multistage EA dynamics.}

\section{Numerical implementation of the Lindblad dynamics}
\label{Appendix:B}

The numerical calculations presented in this work combine time-domain integration of the Lindblad master equation with a complementary spectral analysis of the Liouvillian superoperator. The former determines the dissipative evolution of the density operator, whereas the latter resolves the relaxation dynamics into individual Liouvillian decay modes and provides the basis for the mode-filtering analysis discussed in the main text.

The cavity field is represented in a truncated Fock basis with photon cutoff $n_p$, whereas the atomic ensemble is restricted to the permutation-symmetric Dicke manifold with total spin $J=N/2$. This permutation-symmetric representation reduces the Hilbert-space dimension to
\begin{equation}
D=(n_p+1)(N+1),
\end{equation}
while remaining invariant under the Lindblad dynamics generated by both cavity photon loss and collective atomic decay. Consequently, the complete dissipative dynamics is captured within the permutation-symmetric subspace.

The dissipative evolution is governed by the Lindblad master equation,
\begin{equation}
\frac{d\rho}{dt}
=-i[H,\rho]+\sum_{\mu}\left(L_\mu\rho L_\mu^\dagger-\frac12\{L_\mu^\dagger L_\mu,\rho\}
\right),   
\end{equation}
whose numerical integration yields the time-dependent density matrix $\rho(t)$. At each stage of the evolution, the parity-projected density matrix,
\begin{equation}
\rho_P(t)
=\sum_{p=\pm1}\Pi_p\rho(t)\Pi_p,
\end{equation}
is constructed, and the entanglement asymmetry is evaluated according to Eq.~\ref{eq:EA_Renyi2}.

To elucidate the relaxation dynamics from a spectral perspective, the density operator is represented in Liouville space through the vectorization map,
\begin{equation}
|\rho\rangle\rangle=\mathrm{vec}(\rho), 
\end{equation}
under which the Lindblad master equation assumes the linear form
\begin{equation}
\frac{d}{dt}
|\rho\rangle\rangle
=\mathcal L|\rho\rangle\rangle,
\end{equation}
with the Liouvillian superoperator
\begin{align}
\mathcal L
=&
-i
\left(
I\otimes H
-
H^{\mathrm T}\otimes I
\right)\nonumber\\
&+
\sum_\mu
\left[
L_\mu^{*}\otimes L_\mu
-
\frac12
I\otimes L_\mu^\dagger L_\mu
-
\frac12
(L_\mu^\dagger L_\mu)^{\mathrm T}\otimes I
\right].   
\end{align}

{\color{black}
The Liouvillian constitutes a generally non-Hermitian generator acting on the operator space. Using the Hilbert--Schmidt inner product, its right and left eigenoperators are defined by
\begin{align}
\mathcal L[R_k]&=\mu_k R_k,\\
\mathcal L^\dagger[L_k]&=\mu_k^*L_k,\\
\mathrm{Tr}(L_k^\dagger R_j)&=\delta_{kj},\qquad
\textcolor{black}{\|R_k\|_{\rm HS}=1}.
\label{eq:biorthogonal_convention}
\end{align}

The corresponding spectral representation of the density operator is
\begin{equation}
\rho(t)=\rho_{\rm steady}+
\sum_{k\ge2}c_ke^{\mu_k t}R_k,
\end{equation}
where the expansion coefficients are
\begin{equation}
c_k
=\mathrm{Tr}(L_k^\dagger\rho_0).
\end{equation}
\textcolor{black}{With the above normalization,} these coefficients quantify the contribution of individual Liouvillian decay modes to the relaxation dynamics discussed in the main text.
}

Finally, numerical convergence was established by systematically increasing the cavity photon cutoff. The entanglement-asymmetry dynamics, the low-lying Liouvillian spectrum, and the corresponding mode amplitudes remain unchanged within numerical precision.

\section{Linear-fluctuation origin of the slow parity-odd mode}
\label{app:linear_fluctuation}

We derive the linear-fluctuation structure of the slow
parity-odd Liouvillian mode in the normal phase of the
dissipative Dicke model. This provides a microscopic
interpretation of the approximately linear coherent-state
overlap and the interference-induced suppression obtained
for cat-product initial states.

We consider the Dicke Hamiltonian with cavity photon loss,
\begin{equation}
\begin{aligned}
H
&=
\omega a^\dagger a+\omega_0J_z
+\frac{\lambda}{\sqrt N}
(a+a^\dagger)(J_++J_-),
\\
\mathcal L[\rho]
&=
-i[H,\rho]
+\kappa\left(
a\rho a^\dagger
-\frac{1}{2}\{a^\dagger a,\rho\}
\right).
\end{aligned}
\label{eq:app_model}
\end{equation}
Within the permutation-symmetric sector, $J=N/2$, the
Holstein--Primakoff representation gives
$J_z=b^\dagger b-N/2$. For a small spin-excitation fraction,
$\langle b^\dagger b\rangle/N\ll1$, the ladder operators
reduce to $J_+\simeq\sqrt N\,b^\dagger$ and
$J_-\simeq\sqrt N\,b$. Retaining the quadratic terms and
omitting an additive constant yields
\begin{equation}
H_{\rm lin}
=
\omega a^\dagger a+\omega_0b^\dagger b
+\lambda(a+a^\dagger)(b+b^\dagger).
\label{eq:H_linear}
\end{equation}
The normal-phase fluctuations are therefore described by
two coupled bosonic modes. This expansion assumes small
fluctuations around the polarized normal state; in the
superradiant phase, a Gaussian description instead requires
an expansion around a displaced mean-field solution.

The adjoint Liouvillian generated by $H_{\rm lin}$ closes
on the linear operator subspace
$\operatorname{span}\{a,a^\dagger,b,b^\dagger\}$.
To express this closure, we introduce
$\mathbf X=(x_a,p_a,x_b,p_b)^{\rm T}$, where
$x_a=(a+a^\dagger)/\sqrt2$ and
$p_a=(a-a^\dagger)/(i\sqrt2)$, with analogous definitions
for the atomic mode. Its adjoint evolution is
$\mathcal L^\dagger[\mathbf X]=A\mathbf X$, with
\begin{equation}
A=
\begin{pmatrix}
-\kappa/2 & \omega & 0 & 0
\\
-\omega & -\kappa/2 & -2\lambda & 0
\\
0 & 0 & 0 & \omega_0
\\
-2\lambda & 0 & -\omega_0 & 0
\end{pmatrix}.
\label{eq:drift_matrix}
\end{equation}

The elementary relaxation rapidities are the eigenvalues
of $A$. The characteristic equation $\det(sI-A)=0$ gives
\begin{equation}
\left[
\left(s+\frac{\kappa}{2}\right)^2+\omega^2
\right]
\left(s^2+\omega_0^2\right)
-4\lambda^2\omega\omega_0=0.
\label{eq:rapidity_equation}
\end{equation}
The normal-phase instability occurs when a slow rapidity
reaches $s=0$. Substituting this condition into
Eq.~(\ref{eq:rapidity_equation}) yields
\begin{equation}
\left(\omega^2+\frac{\kappa^2}{4}\right)\omega_0^2
=
4\lambda_c^2\omega\omega_0,
\qquad
\lambda_c
=
\frac{1}{2}
\sqrt{
\frac{(\omega^2+\kappa^2/4)\omega_0}{\omega}
}.
\label{eq:lambda_c_linear}
\end{equation}
Thus, the same linear-fluctuation theory identifies both
the slow relaxation channel and its softening at the
normal-to-superradiant transition.

A left eigenvector satisfying
$\mathbf w^{\rm T}A=s\mathbf w^{\rm T}$ defines an
adjoint eigenoperator $L_s=\mathbf w^{\rm T}\mathbf X$.
In terms of the bosonic operators, this takes the form
\begin{equation}
\begin{aligned}
L_s
&=
u_a a+v_a a^\dagger+u_b b+v_b b^\dagger,
\\
\mathcal L^\dagger[L_s]
&=
sL_s.
\end{aligned}
\label{eq:linear_eigenoperator}
\end{equation}
Under Dicke parity, both $a$ and $b$ change sign, so
$PL_sP^\dagger=-L_s$. These elementary fluctuation modes
therefore belong to the parity-odd sector. Within the stable
Gaussian theory, higher-order relaxation modes involve
sums of elementary rapidities, and the least-damped
first-order modes provide the leading odd relaxation
channels. In the convention
$\mathcal L^\dagger[L_2]=\mu_2^*L_2$, the corresponding
rapidity is $s=\mu_2^*$.

For the initial states considered in the main text, the
atomic ensemble is prepared in $|J,-J\rangle$, corresponding
to the Holstein--Primakoff vacuum $|0_b\rangle$.
Projecting the slow left eigenoperator onto this vacuum
gives the effective cavity profile
$\ell_2\equiv\langle0_b|L_2|0_b\rangle
\simeq u_a a+v_a a^\dagger$.
The atomic terms vanish because
$\langle b\rangle=\langle b^\dagger\rangle=0$.
Using the biorthogonal convention
$\operatorname{Tr}(L_k^\dagger R_j)=\delta_{kj}$,
with $\|R_k\|_{\rm HS}=1$, the mode coefficient is
$c_k=\operatorname{Tr}(L_k^\dagger\rho_0)$.
For the coherent-product state
$|\alpha\rangle\otimes|0_b\rangle$ with real $\alpha$,
this gives
\begin{equation}
c_2^{\rm coh}
=
\langle\alpha|\ell_2^\dagger|\alpha\rangle
\simeq
(u_a^*+v_a^*)\alpha.
\label{eq:c2_coh_linear}
\end{equation}
The approximately linear dependence of
$|c_2^{\rm coh}|$ on $\alpha$ thus reflects the direct
excitation of the slow light--matter fluctuation by
the cavity coherent displacement.

For the cat-product initial state, we write
\begin{equation}
\begin{aligned}
|\psi_{\rm cat}\rangle
&=
\mathcal N
\left(
|\alpha\rangle+e^{i\theta}|-\alpha\rangle
\right)\otimes|0_b\rangle,
\\
\mathcal N^{-2}
&=
2\left[1+\cos\theta\,e^{-2\alpha^2}\right],
\end{aligned}
\label{eq:cat_general_app}
\end{equation}
where $\alpha>0$ is real. Since $\ell_2^\dagger$ is
parity odd, its diagonal matrix elements on the two
coherent branches have opposite signs,
$\langle-\alpha|\ell_2^\dagger|-\alpha\rangle
=-\langle\alpha|\ell_2^\dagger|\alpha\rangle$.
Their contributions therefore cancel, leaving only
the inter-branch coherences. Using
$\langle\alpha|-\alpha\rangle=e^{-2\alpha^2}$,
one obtains
$\langle\alpha|\ell_2^\dagger|-\alpha\rangle
\simeq(u_a^*-v_a^*)\alpha e^{-2\alpha^2}$.
The reverse matrix element has the opposite sign, so
the two cross terms combine through
$e^{i\theta}-e^{-i\theta}=2i\sin\theta$. Consequently,
\begin{equation}
|c_2^{\rm cat}|
\simeq
|u_a^*-v_a^*|
\frac{
|\sin\theta|\,\alpha e^{-2\alpha^2}
}{
1+\cos\theta\,e^{-2\alpha^2}
}.
\label{eq:c2_cat_linear_theory}
\end{equation}

The two initial-state families therefore excite the same
slow parity-odd channel in different ways.
A coherent-product state couples directly through its
displacement, giving $|c_2^{\rm coh}|\propto\alpha$.
For a cat-product state, cancellation between the diagonal
branch contributions leaves a residual overlap controlled
by the inter-branch coherence. As the coherent components
become well separated, the factor $e^{-2\alpha^2}$
suppresses this residual contribution.
This interference mechanism provides the normal-phase
linear-fluctuation interpretation of initial-state
mode filtering: suppressing the slow channel can allow
the remaining, faster contributions to govern relaxation
over an extended time interval.
\end{document}